# A Novel Variable Step Size NLMS Algorithm Based on the Power Estimate of the System Noise

Yi Yu, and Haiquan Zhao
School of Electrical Engineering, Southwest Jiaotong University, Chengdu, China
E-mail: yuyi_xyuan@163.com, hqzhao@home.swjtu.edu.cn

*Abstract*—To overcome the tradeoff of the conventional normalized least mean square (NLMS) algorithm between fast convergence rate and low steady-state misalignment, this paper proposes a variable step size (VSS) NLMS algorithm by devising a new strategy to update the step size. In this strategy, the input signal power and the cross-correlation between the input signal and the error signal are used to estimate the true tracking error power, reducing the effect of the system noise on the algorithm performance. Moreover, the steady-state performances of the algorithm are provided for Gaussian white input signal and are verified by simulations. Finally, simulation results in the context of the system identification and acoustic echo cancellation (AEC) have demonstrated that the proposed algorithm has lower steady-state misalignment than other VSS algorithms.

## I. INTRODUCTION

Over the past decades, adaptive filtering algorithms have been widely applied in many signal processing applications such as system identification, active noise cancellation (ANC), channel equalization, acoustic echo cancellation (AEC) and so on [1-2], therein the normalized least mean square (NLMS) algorithm is the most simple and popular algorithm. In the NLMS algorithm, however, the choice of the step size must take into account a compromise between fast convergence rate and low steady-state error.

Aiming to solve this problem, several variable step size (VSS) NLMS algorithms have been proposed [3-14]. For these VSS schemes, the fundamental idea is that the step size has a larger value in the early stage of adaptive processing to speed up convergence; then its value is decreased gradually as the iteration goes on; when the algorithm converges to the steady-state, the value of the step size is small to yield a low steady-state error. Importantly, different VSS schemes can yield different the algorithm performance. In [3], a simple VSS method was presented by using the instantaneous squared error to adjust the step size. However, the update of the step size is subject to the disturbance of the system noise, especially in the low signal-to-noise rate (SNR) environments. To eliminate this interference, robust VSS algorithm [4] and noise resilient VSS algorithm [5] were proposed. Whereas, the tracking capability of the algorithm in [4] is very poor in non-stationary environments and the algorithm in [5] requires knowing the input signal power in advance. Although Mandic proposed a variable regularization parameter NLMS algorithm to improve the convergence rate, essentially it can also be regarded as a VSS NLMS algorithm [6]. Also, the adjusting of the regularization parameter is not robust enough to cope with the varying of the unknown system. Based on the weighting average of the cross-correlation between the output error and the input vector, some VSS schemes were designed [7-9]. Nevertheless, the performance of the algorithm in [7] is deteriorated as the SNR decreases, and Hwang's algorithm is instable for the time-varying unknown system [9]. In [10] and [11], two nonparametric VSS algorithms were developed, i.e., NPVSS and NVS-NLMS, respectively, which provide good convergence performance with the prior knowledge of the system noise power. As an improvement of the NPVSS, the NEW-NPVSS algorithm [12] employs the estimated system noise power to adjust the step size at the expense of the computational complexity and steady-state error. Subsequently, two VSS NLMS algorithms [13-14] were proposed based on the approach used to estimate the system noise power.

Inspired by the method presented in [15], a new VSS NLMS algorithm is proposed to address the tradeoff problem in the conventional NLMS. Also, the true tracking error power is estimated using the method in [12] to update the step size. Although the algorithm has been reported in [16], this paper is more comprehensive. Namely, the main contributions of this paper are listed as follows:

1) The convergence explanation of the algorithm is more intuitive via using Fig. 2.
2) The steady-state performances of the algorithm are analyzed in terms of the excess mean-square-error (EMSE) and mean-square-deviation (MSD) and verified by simulations. The similar analysis has been performed in [16], but the results are questionable due to an unreasonable approximation to simplify (15) in [16].
3) The computational complexity and parameters' selection of the algorithm are discussed.
4) An application of the algorithm such as AEC is conducted.

## II. ALGORITHMS

In this section, the conventional NLMS algorithm is firstly described, and then a new VSS NLMS algorithm is proposed.

### A. NLMS

Consider a system identification problem, as shown in Fig. 1, where $\mathbf{w}_o$ denotes the unknown *M*-dimensional column vector, i.e.,

$$\mathbf{w}_o = \begin{bmatrix} \mathbf{w}_{o0} & \mathbf{w}_{o1} & \cdots & \mathbf{w}_{oM-1} \end{bmatrix}^T, \quad (1)$$

that we want to estimate by using an adaptive filter $\mathbf{w}(n)$ and the superscript $T$ indicates transposition.

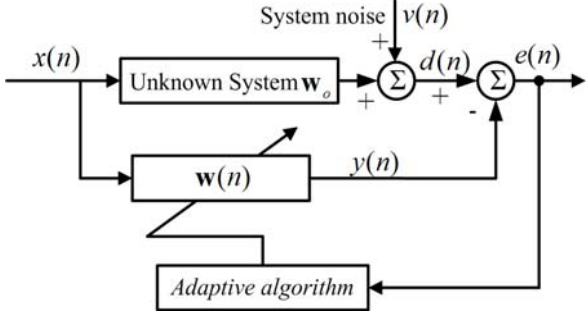

Figure 1. Structure of adaptive system identification.

Then, the desired signal $d(n)$ of the adaptive filter can be given by the formula

$$d(n) = \mathbf{w}_o^T \mathbf{x}(n) + v(n) \quad (2)$$

where $n$ is the time index, $\mathbf{x}(n)$ is the input vector, i.e.,

$$\mathbf{x}(n) = [x(n), x(n-1), \ldots, x(n-M+1)]^T, \quad (3)$$

and $v(n)$ is the system noise with zero-mean and variance $\sigma_v^2$. Thus, the weight vector of the conventional NLMS algorithm can be updated as

$$\mathbf{w}(n+1) = \mathbf{w}(n) + \mu \frac{e(n)\mathbf{x}(n)}{\delta + \mathbf{x}^T(n)\mathbf{x}(n)} \quad (4)$$

where $1 < \mu < 2$ is the fixed step size and the output error $e(n)$ is given by

$$e(n) = d(n) - y(n) = d(n) - \mathbf{w}^T(n)\mathbf{x}(n) \quad (5)$$

and $\delta > 0$ is a small regularization parameter to avoid division by zero. Interestingly, the step size controls the steady-state error and convergence rate of the algorithm, i.e., a small step size yields a small steady-state error but slows convergence, vice versa.

B. *Proposed VSS NLMS*

In order to overcome the inherent tradeoff of the conventional NLMS algorithm, inspired by variable forgetting factor idea in [15], we obtain a new, easy to implement, NLMS algorithm with a time-varying step size $\mu(n)$. The VSS mechanism is given by

$$\mu(n) = \mu_{\max} + (\mu_{\min} - \mu_{\max})\exp^{-\beta \sigma_c^2(n)} \quad (6)$$

where $\exp(\cdot)$ denotes the exponential function; $\mu_{\min}$ and $\mu_{\max}$ are the minimum and maximum step sizes to provide a minimum level of tracking capability of the algorithm and to ensure stability, respectively; the parameter $\beta$ is a positive number which adds the flexibility of designing algorithm and will be investigated in the next section. In (6), $\sigma_c^2(n)$ represents the power of the true tracking error $c(n)$ to eliminate the effect of the system noise on the step size, wherein $c(n)$ can be obtained by subtracting the system noise $v(n)$ from the output error $e(n)$, i.e.,

$$c(n) = e(n) - v(n) = \mathbf{x}^T(n)[\mathbf{w}_o - \mathbf{w}(n)]. \quad (7)$$

Although $\sigma_c^2(n)$ is unknown and time-varying due to the fact that $\mathbf{w}_o$ is unknown, it can be effectively estimated by the following expression [12], i.e.,

$$\sigma_c^2(n) = E\left[c^2(n)\right] \approx \frac{\gamma_{ex}^T(n)\gamma_{ex}(n)}{\sigma_x^2(n)} \quad (8)$$

where $E[\cdot]$ denotes the mathematical expectation, $\sigma_x^2(n) = E\left[x^2(n)\right]$ is the input signal power. In (8), $\gamma_{ex}(n)$ denotes the cross-correlation between the input vector $x(n)$ and the output error $e(n)$, and is defined by

$$\begin{aligned}\gamma_{ex}(n) &= E[e(n)\mathbf{x}(n)] \\ &= E\left\{\mathbf{x}(n)\mathbf{x}^T(n)[\mathbf{w}_o - \mathbf{w}(n)]\right\} + E[\mathbf{x}(n)v(n)] \quad (9) \\ &= \mathbf{R}_{xx}(n)E[\Delta\mathbf{w}(n)]\end{aligned}$$

where $\mathbf{R}_{xx}(n) = E\left[\mathbf{x}(n)\mathbf{x}^T(n)\right]$ is the autocorrelation matrix of the input vector, and $\Delta\mathbf{w}(n) = \mathbf{w}_o - \mathbf{w}(n)$. It is worthy to note that equation (9) derives from a commonly used assumption that the system noise $v(n)$ is independent of the input signal $x(n)$. In (8), the values of $\sigma_x^2(n)$ and $\gamma_{ex}(n)$ are exact, but not available in practice. In general, they can be approximated by using the weighting average [12], i.e.,

$$\sigma_x^2(n) = \alpha\sigma_x^2(n-1) + (1-\alpha)x^2(n) \quad (10)$$
$$\gamma_{ex}(n) = \alpha\gamma_{ex}(n-1) + (1-\alpha)e(n)\mathbf{x}(n) \quad (11)$$

where $\alpha$ is the weighting factor. In addition, to avoid that the value of the denominator in (8) is zero when there is no excitation, i.e., $\sigma_x^2(n) = 0$ (which is often encountered in AEC), a small positive number $\rho$ should be added and then (8) is rewritten as

$$\sigma_c^2(n) \approx \frac{\gamma_{ex}^T(n)\gamma_{ex}(n)}{\rho + \sigma_x^2(n)}. \quad (12)$$

III. PERFORMANCE ANALYSES

In the section, the performance analysis of the proposed algorithm is performed in terms of the convergence explanation, steady-state EMSE and MSD, computational complexity and parameters' selection.

A. *Convergence explanation*

Fig. 2 depicts the functional relationship between the step size $\mu(n)$ and $\sigma_c^2(n)$ which is given in (6). From Fig. 2, we can obtain the following conclusions. In the initial stage of the adaptive processing or when the unknown vector $\mathbf{w}_o$ changes suddenly, $\sigma_c^2(n)$ has a larger value due to the system mismatch. Thus, $\exp^{-\beta\sigma_c^2(n)}$ is very small such that the proposed algorithm gets a large step size (close to $\mu_{\max}$) from (6) to improve the convergence rate. As the algorithm goes on, $\sigma_c^2(n)$ becomes gradually small such that $\exp^{-\beta\sigma_c^2(n)}$ becomes increasingly close to one, and thereby the step size $\mu(n)$ gets small. When the algorithm has converged to the steady-state, $\sigma_c^2(n)$ is pretty close to zero, leading to a small step size (close to $\mu_{\min}$) to obtain a low steady-state error. In addition, the true tracking error power $\sigma_c^2(n)$ is used to update the step size, so the step size is insensitive to the system noise.

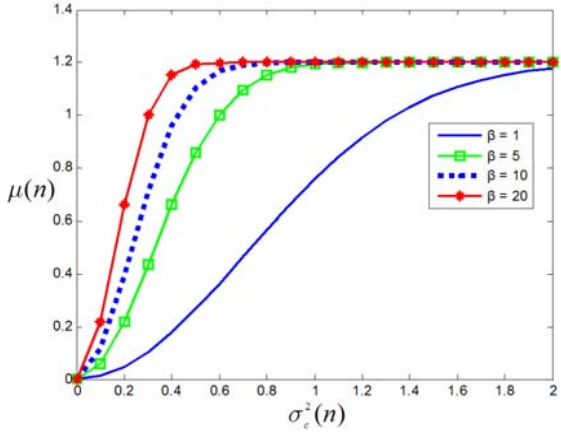

Figure 2. The relation curve between the step size $\mu(n)$ and $\sigma_c^2(n)$ for four different values of $\beta$

## B. Steady-state EMSE and MSD

In order to derive the steady-state EMSE and MSD, we firstly employ the following assumptions: 1) the proposed algorithm has converged to the steady-state; 2) the misadjustment $\xi$ is small and the regularization parameter $\delta$ is much smaller than $\mathbf{x}^T(n)\mathbf{x}(n)$. Let $\xi_{\min}$ denote the minimum mean-square-error (Wiener error), i.e., $\xi_{\min} = \sigma_v^2$, and $\xi_{ex}$ represent the steady-state EMSE. According to the results in [1] and [13], we can obtain the misadjustment of the conventional NLMS with a constant step size $\mu$, which is expressed as

$$\zeta = \frac{\xi_{ex}}{\xi_{\min}} \approx \frac{\mu}{2-\mu}. \tag{13}$$

In the steady-state stage, i.e., $n \to \infty$, and taking the expectation of both sides of (6), we can obtain

$$E[\mu(\infty)] = \mu_{\max} + (\mu_{\min} - \mu_{\max})E\left[\exp^{-\beta\sigma_c^2(\infty)}\right]. \tag{14}$$

Because the $\beta\sigma_c^2(\infty)$ is very small in the steady-state stage, we can use the first-order Taylor series, i.e., $\exp^{-\beta\sigma_c^2(\infty)} \approx 1 - \beta\sigma_c^2(\infty)$, to simplify the exponential term at the right side of (14). Thus, (14) is modified as

$$E[\mu(\infty)] = \mu_{\min} + (\mu_{\max} - \mu_{\min})\beta E[\sigma_c^2(\infty)]. \tag{15}$$

For the analysis tractable, assuming that the input signal $x(n)$ is zero-mean Gaussian white, and the excess error in the steady-state is much smaller than the system noise, i.e., $e(n) \approx v(n)$. Based on these assumptions, we can thus use equation (10) given in [9], i.e.,

$$E\left[\boldsymbol{\gamma}^T(\infty)\boldsymbol{\gamma}(\infty)\right] \approx \frac{1-\alpha}{1+\alpha}M\sigma_x^2\sigma_v^2. \tag{16}$$

By observing (10), it can be rewritten as

$$\sigma_x^2(n) = (1-\alpha)\sum_{i=0}^{n-1}\alpha^i x^2(n-i). \tag{17}$$

Assuming that the weighting factor $\alpha$ is close to one such that the ensemble average of $\sigma_x^2(n)$ can be approximated by its time average. Thus, taking the expectation of (17), we have

$$E\left[\sigma_x^2(n)\right] \approx E\left[x^2(n)\right] \approx \sigma_x^2. \tag{18}$$

Substituting (12), (16) and (18) into (15) yields

$$E[\mu(\infty)] = \mu_{\min} + (\mu_{\max} - \mu_{\min})\beta M\frac{1-\alpha}{1+\alpha}\sigma_v^2. \tag{19}$$

To derive (19), a reasonable approximation that the expected value of the ratio between random variables is approximately equal to the ratio between the expected values of these variables is used, i.e., $E[X/Y] \approx E[X]/E[Y]$.

Then, using $E[\mu(\infty)]$ instead of the constant $\mu$ in (13), we can obtain the steady-state EMSE of the proposed algorithm, i.e.,

$$\xi_{ex} \approx$$

$$\frac{\mu_{\min}(1+\alpha) + \beta M(\mu_{\max}-\mu_{\min})(1-\alpha)\sigma_v^2}{(2-\mu_{\min})(1+\alpha) - \beta M(\mu_{\max}-\mu_{\min})(1-\alpha)\sigma_v^2}\sigma_v^2 \tag{20}$$

with

$$\beta < \frac{(2-\mu_{\min})(1+\alpha)}{M(\mu_{\max}-\mu_{\min})(1-\alpha)\sigma_v^2}. \tag{21}$$

Next, we derive the steady-state MSD of the proposed algorithm. The MSD is defined as [1][13]

$$\text{MSD}(n) = E\left\{\|\mathbf{w}(n)-\mathbf{w}_o\|^2\right\}. \tag{22}$$

Similarly, according to the results in [1] and [13], the steady-state MSD of the conventional NLMS algorithm can be computed by

$$\text{MSD}(\infty) \approx \frac{\mu M\sigma_v^2}{(2-\mu)M\sigma_x^2 + 2\delta}. \tag{23}$$

Then, substituting (19) into (23), the steady-state MSD of the proposed algorithm is shown in (24).

## C. Computational complexity

In TABLE I, the computational complexity of the proposed algorithm is compared with that of other VSS algorithms in terms of the total number of multiplications, additions, exponents and square roots. The proposed algorithm with filter length $M$ requires $5M+7$ additions, $5M+10$ multiplications and one exponent for implementing per iteration. The proposed algorithm has larger computational complexity than the NPVSS and NVS-NLMS algorithms, but smaller than the NEW-NPVSS algorithm.

TABLE I. COMPUTATIONAL COMPLEXITY OF VARIOUS VSS ALGORITHMS

| Algorithms | Multiplications | Additions | Exponents | Square-roots |
|---|---|---|---|---|
| NPVSS | $3M+6$ | $3M+6$ | 0 | 2 |
| NVS-NLMS | $3M+6$ | $3M+6$ | 0 | 0 |
| NEW-NPVSS | $5M+16$ | $5M+10$ | 0 | 1 |
| Proposed algorithm | $5M+9$ | $5M+7$ | 1 | 0 |

## D. Parameters' selection

$$\text{MSD}(\infty) \approx \frac{\mu_{\min}M(1+\alpha)\sigma_v^2 + \beta M^2(1-\alpha)(\mu_{\max}-\mu_{\min})\sigma_v^4}{(1+\alpha)\left[(2-\mu_{\min})M\sigma_x^2 + 2\delta\right] - \beta M^2(1-\alpha)(\mu_{\max}-\mu_{\min})\sigma_x^2\sigma_v^2}. \tag{24}$$

In order to better reflect the practicability of the proposed algorithm, the choices of its parameters are discussed in this section.

1) To ensure the stability and fast convergence rate of the algorithm, generally, the maximum step size $\mu_{\max}$ should be chosen in the range of $1 < \mu_{\max} < 2$. The minimum step size $\mu_{\min}$ should be very small for acquiring low steady-state error. After testing, $\mu_{\max}$ and $\mu_{\min}$ are set to 1.2 and 0.001 in this paper, respectively.
2) The regularization parameter $\delta$ can be chosen by the rule given in [17], i.e., $\delta = M\left(1+\sqrt{1+\mathrm{SNR}}\right)\sigma_x^2 / \mathrm{SNR}$.
3) The parameter $\rho$ in (12) is a very small positive number to avoid the division by zero, thus it can easily be selected.
4) To accurately estimate $\sigma_c^2(n)$, the weighting factor $\alpha$ is often selected using $\alpha = 1 - 1/\kappa M$ with $\kappa \geq 2$.
5) A conclusion can be drawn from (20) and (21) that the range of the parameter $\beta$ is dependent on the weighting factor $\alpha$, filter length $M$, variance of the system noise $\sigma_v^2$, maximum step size $\mu_{\max}$ and minimum step size $\mu_{\min}$. It can be seen from (20) that a small value of $\beta$ can lead to a low steady-state EMSE. However, the value also slows convergence, since the step size $\mu(n)$ is quickly decreased in the beginning stage of adaptive processing as shown in Fig. 2. Therefore, in practical applications, we firstly provide a suitable range for $\beta$ using (20) and (21), and then select a proper value of $\beta$ by a heuristic search.

## IV. SIMULATION RESULTS

In this section, the performance of the proposed algorithm is compared with that of the NPVSS, NVS-NLMS, NEW-NPVSS algorithms in the system identification and AEC environments. The normalized misalignment (in dB), $20\log_{10}\left(\|\mathbf{w}_o - \mathbf{w}(n)\|_2 / \|\mathbf{w}_o\|_2\right)$, is used to measure the performance of these algorithms. The initial weight vectors for all algorithms are zero, and other parameters' values are shown in TABLE II. All results are the ensemble average of 100 independent runs except for section 4.2.

TABLE II. LIST OF PARAMETERS' VALUE FOR VARIOUS VSS ALGORITHMS.

| Algorithms | Parameters |
|---|---|
| NPVSS | $\kappa=2$, $\varepsilon=2$ |
| NVS-NLMS | $\kappa=2$, $\varepsilon=2$ |
| NEW-NPVSS | $\kappa=2$, $\varepsilon_{th}=2$ |
| Proposed algorithm | $\mu_{\max}=1.2$, $\mu_{\min}=0.001$, $\kappa=2$, $\beta=20$ |

### A. System identification

The unknown vector $\mathbf{w}_o$ with length of $M = 10$, is randomly generated by using the rand function in MATLAB and normalized by using $\mathbf{w}_o^T \mathbf{w}_o = 1$, and is changed as $-\mathbf{w}_o$ at iteration 1000. The input signal is either a Gaussian white noise with variance $\sigma_x^2 = 1$ or an AR(1) process generated by filtering a Gaussian white noise through a first order auto-regression system with a pole at 0.5.

*1) Effect of varying β*

In order to illustrate the effect of the parameter $\beta$ on the performance of the proposed algorithm, in this experiment, five different values of $\beta$ (i.e., 5, 15, 20, 25, and 40) are used. The input signal is Gaussian white noise, and the system noise power is $\sigma_v^2 = 0.01$.

Fig. 3 shows the misalignment curves of the proposed algorithm for different values of $\beta$. It is clear that the proposed algorithm with small value of $\beta$ exhibits low steady-state misalignment, and it has also a slow convergence rate. Note that, for the value of $\beta$ in the range (15, 25), the proposed algorithm has the advantages of both fast convergence rate and low steady-state error. In addition, equations (20) and (21) can provide some guidance on the choice of the parameter $\beta$.

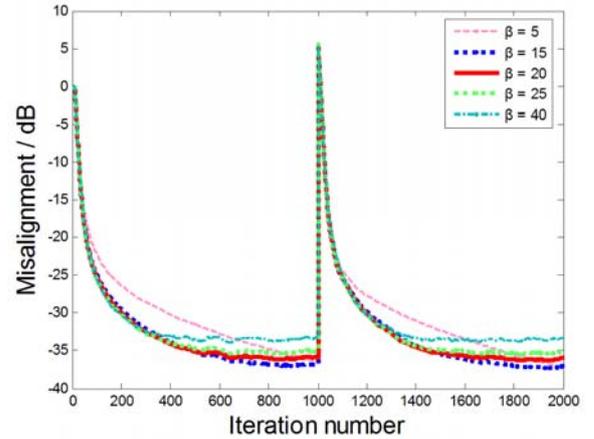

Figure 3. Effect of different values of $\beta$

*2) Comparison with analytical calculation*

This example gives a comparison of the theoretical EMSE in (20) with simulated results. The unknown vector wo is randomly generated. Several cases are considered: 1) input signals $x(n)$: Gaussian white and AR(1) process; and 2) length $M$ = 64 and 128; 3) system noise levels: $\sigma_v^2 = 0.01$ ($\beta=20$) and $\sigma_v^2 = 0.09$ ($\beta=5$). As a result, the experimental values of the steady-state EMSE and its theoretical values calculated by (20) are shown in TABLE III. Clearly, there is only a slight difference between theoretical values and experimental values due to using some assumptions and approximations for deriving (20). Therefore, (20) can be used to predict the steady-state EMSE of the proposed algorithm. Here, a verification of the theoretical MSD given by (24) is omitted, due to a simple relation between EMSE and MSD for white input signals, i.e., $\mathrm{EMSE}(n) = \sigma_x^2 \mathrm{MSD}(n)$.

*3) Gaussian white input*

In this case, the input signal is Gaussian white noise. Fig. 4 describes the misalignment curves of the algorithms for system noise power $\sigma_v^2 = 0.01$. Fig. 5 describes the misalignment curves of the algorithms in the scenario that the system noise power changes from 0.01 to 0.09 at iteration 1000.

TABLE III. THEORETICAL AND EXPERIMENTAL VALUES OF THE STEADY-STATE EMSE.

| Input signals | $M$ | Noise levels $\sigma_v^2$ | Theoretical values $\xi_{ex}$ | Experimental values $\widehat{\xi_{ex}}$ | Relative error $\left|\frac{\xi_{ex} - \widehat{\xi_{ex}}}{\xi_{ex}}\right|$ |
|---|---|---|---|---|---|
| White Gaussian | 64 | 0.01 | $3.1558\times 10^{-4}$ | $3.2384\times 10^{-4}$ | 2.617% |
| White Gaussian | 128 | 0.01 | $3.1495\times 10^{-4}$ | $3.1704\times 10^{-4}$ | 0.664% |
| AR(1) process | 64 | 0.01 | $3.1558\times 10^{-4}$ | $3.2491\times 10^{-4}$ | 2.956% |
| AR(1) process | 128 | 0.01 | $3.1495\times 10^{-4}$ | $3.3004\times 10^{-4}$ | 4.791% |
| White Gaussian | 64 | 0.09 | $6.5881\times 10^{-3}$ | $6.3221\times 10^{-3}$ | 4.038% |
| White Gaussian | 128 | 0.09 | $6.5774\times 10^{-3}$ | $5.8086\times 10^{-3}$ | 11.689% |
| AR(1) process | 64 | 0.09 | $6.5881\times 10^{-3}$ | $6.0021\times 10^{-3}$ | 8.894% |
| AR(1) process | 128 | 0.09 | $6.5774\times 10^{-3}$ | $5.8274\times 10^{-3}$ | 11.402% |

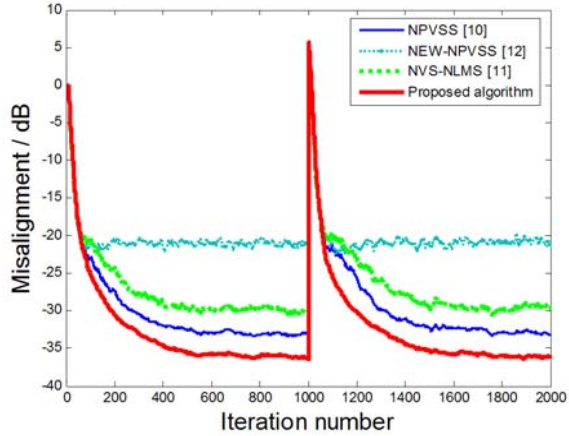

Figure 4. Misalignment of various VSS-NLMS algorithms using Gaussian white input for the system noise power $\sigma_v^2 = 0.01$.

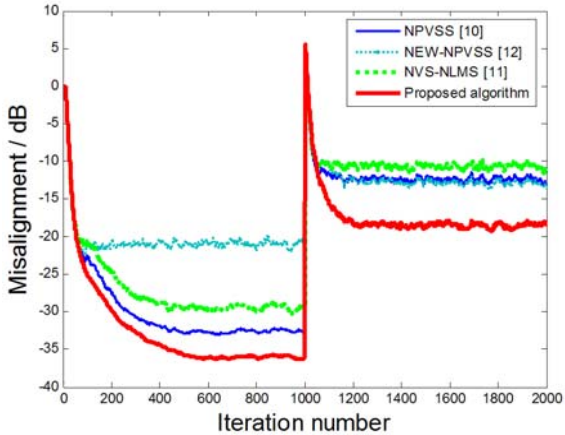

Figure 5. Misalignment of various VSS-NLMS algorithms using Gaussian white input for varying system noise power.

In Fig. 4, both NPVSS and proposed algorithms have lower steady-state misalignment than the NEW-NPVSS and NVS-NLMS algorithms. Also, the steady-state misalignment of the NPVSS is greater than that of the NEW-NPVSS when the system noise power increases from 0.01 to 0.09 (see Fig. 5). More importantly, only the proposed algorithm has the best performance in the steady-state misalignment, even if the system noise power increases. The reason behind the results is that the NPVSS and NVS-NLMS depend on the prior knowledge of the system noise power for achieving better performance, and which are suitable for the system environment with noise power unchanged. However, the proposed algorithm does not need to know the system noise power, since the true tracking error power $\sigma_c^2(n)$ used to update the step size can be effectively estimated. Although the NEW-NPVSS algorithm does also not require the system noise power, it increases the computational cost as well as steady-state misalignment.

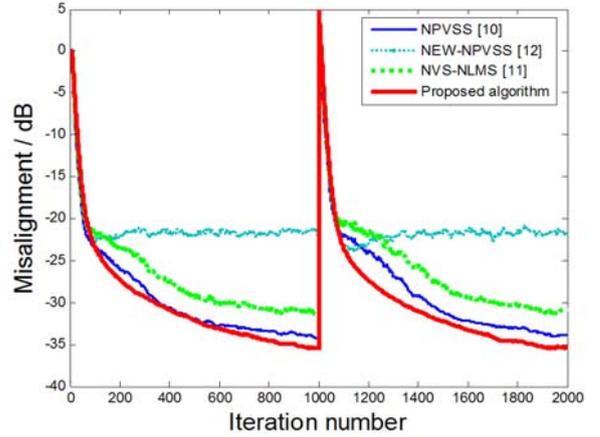

Figure 6. Misalignment of various VSS-NLMS algorithms using AR(1) input for the system noise power $\sigma_v^2 = 0.01$.

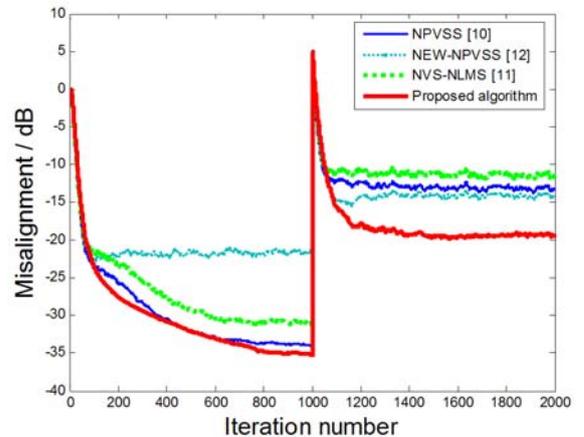

Figure 7. Misalignment of various VSS-NLMS algorithms using AR(1) input for varying system noise power.

*4) AR(1) input*

This case considers that the input signal is an AR(1) process. The misalignment curves of the algorithms are shown in Fig. 6 (system noise power $\sigma_v^2 = 0.01$) and Fig. 7 (varying noise power which is the same as Fig. 5). From Figs. 6 and 7, we can obtain consistent conclusions with Figs. 4 and 5. Namely, even if the input signal is colored (AR(1) process), the proposed algorithm still has the smallest steady-state misalignment among these algorithms. Although the NPVSS algorithm is more outstanding than the algorithms in [12] and [13], its performance is sensitive to the varying of the system noise power.

*B. AEC*

In this section, we consider an AEC application, where the main objective of adaptive filter is to estimate the impulse response of the true acoustic echo path. Here, the true impulse response $\mathbf{w}_o$ has $M = 512$ coefficients (the sampling rate is 8 kHz) and is changed as $-\mathbf{w}_o$ at iteration 10000, and the input signal is a speech signal. Fig. 8 shows the operated results of the algorithms for the system noise power $\sigma_v^2 = 0.01$. As can be seen from Fig. 8, the NPVSS is superior to the NVS-NLMS and NEW-NPVSS in terms of the steady-state misalignment under the same convergence rate, while the proposed algorithm has smaller steady-state misalignment than the NPVSS.

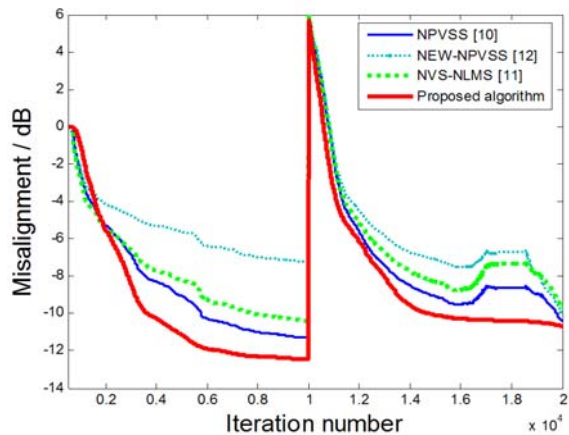

Figure 8. Misalignment of various VSS-NLMS algorithms for AEC.

## V. CONCLUSIONS

In this work, a novel VSS NLMS algorithm is proposed by utilizing the system input power and the cross-correlation between input vector and error signal to implement VSS scheme. The update of the step size is insensitive to the system noise power. In addition, the performances of the algorithm are analyzed such as the convergence explanation, steady-state EMSE and MSD, computational complexity, and parameters' selection. Simulation results in the context of system identification and AEC have demonstrated that the proposed algorithm has lower steady-state misalignment than the NPVSS, NVS-NLMS and NEW-NPVSS algorithms.

ACKNOWLEDGMENT

This work was partially supported by National Science Foundation of P.R. China (Grant: 61271340 and 61071183), the Sichuan Provincial Youth Science and Technology Fund (Grant: 2012JQ0046), and the Fundamental Research Funds for the Central Universities (Grant: SWJTU12CX026).

REFERENCES

[1] S.Haykin, "*Adaptive filter theory (4th ed.)*," Englewood Cliffs: Prentice-Hall, 2001.
[2] Y. Huang, J. Benesty, and J. Chen, "*Acoustic MIMO Signal Processing*," Boston, MA: Springer, 2006.
[3] R. H. Kwong, and E. W. Johnston, "A variable step size LMS algorithm," *IEEE Transactions on Signal Processing*, vol. 40, no. 7, pp. 1633–1642, 1992.
[4] T. Aboulnasr, and K. Mayyas, "A robust variable step size LMS-type algorithm: Analysis and simulations," *IEEE Transactions on Signal Processing*, vol. 45, np. 3, pp. 631–639, 1997.
[5] M. H. Costa, and J. C. M. Bermudez, "A noise resilient variable step size LMS algorithm," *Signal Processing*, vol. 88, no. 3, pp. 733-748, 2008.
[6] D. P. Mandic, "A generalized normalized gradient descent algorithm," *IEEE Signal Processing Letters*, vol. 11, no. 2, pp. 115-118, 2004.
[7] H. C. Shin, A. H. Sayed, and W. J. Song, "Variable step size NLMS and affine projection algorithms," *IEEE Signal Processing Letters*, vol. 11, no. 2, pp. 132–135, 2004.
[8] Y. Zhang, N. Li, J. A. Chambers, Y. Hao, "New gradient-based variable step size LMS algorithms," *EURASIP Journal Advances in Signal Processing*, vol. 2008, no. 7, pp. 1-9, 2008.
[9] E. Hwang, Y. Li, "Variable Step size LMS Algorithm With a Gradient-Based Weighted Average," *IEEE Signal Processing Letters*, vol. 16, no. 12, pp. 1043-1046, 2009.
[10] J. Benesty, H. Rey, L. R. Vega, S. Tressens, "A nonparametric VSS NLMS algorithm," *IEEE Signal Processing Letters*, vol. 13, no. 10, pp. 581–584, 2006.
[11] J. Liu, X. Yu, H. Li, "A nonparametric variable step size NLMS algorithm for transversal filters," *Applied Mathematics and Computation*, vol. 217, pp. 7365–7371, 2011.
[12] M. A. Iqbal, S. L. Grant, "Novel variable step size NLMS algorithms for echo cancellation," *In Proc. IEEE Int. Conf. Acoust., Speech, Signal Processing*, pp. 241–244, 2008.
[13] H. Huang, J.Lee, "A New Variable Step size NLMS Algorithm and Its Performance Analysis," *IEEE Transactions on Signal Processing*, vol. 60, no. 4, pp. 2055–2060, 2012.
[14] Y. Yu, H.Zhao, "An improved variable step size NLMS algorithm based on a Versiera function," *In Proc. IEEE International Conference on Signal Processing, Communication and Computing*, pp. 1–4, 2013.
[15] D. J. Park, B. E. Jun, J. H. Kim, "Fast tracking RLS algorithm using novel variable forgetting factor with unity zone," *Electronics Letters*, vol. 27, no. 23, pp. 2150–2151, 1991.
[16] H. Zhao, Y.Yu, "Novel adaptive VSS-NLMS algorithm for system identification," *In Proc. Fourth International Conference on Intelligent Control and Information Processing*, pp. 760–764, 2013.
[17] J. Benesty, C. Paleologu, S. Ciochina, "On Regularization in Adaptive Filtering," IEEE Transactions on Audio, Speech, and Language Processing, vol. 19, no. 6, pp. 1734–1742, 2011.